\documentclass[12pt]{article}

\usepackage{amsmath,amsbsy,amsfonts,amssymb,amsthm,dsfont,fullpage,units,bm}
\usepackage{mathrsfs,bbm,empheq}
\usepackage[dvips]{graphicx}
\usepackage{subcaption}
\usepackage{algorithm,algorithmic,mathtools}
\usepackage{color,cases}
\usepackage{tikz}
\usepackage{multirow}
\usepackage{hyperref}
\usepackage{comment}
\usepackage{authblk}

\hypersetup{
colorlinks = true,
citecolor = {blue},
citebordercolor = {white}
}

\usepackage[top=1.in, bottom=1in, left=1.1in, right=1.1in]{geometry}

\newcommand{\bi}{\begin{itemize}}
\newcommand{\ei}{\end{itemize}}

\theoremstyle{definition}

\newcommand{\R}{\mathbb{R}}

\newcommand{\EE}{\mathbb{E}}
\newcommand{\E}{\mathbb{E}}

\newcommand{\bx}{\mathbf{x}}

\newcommand{\bw}{\bm{w}}

\newcommand{\bW}{\bm{W}}

\newcommand{\bU}{\bm{U}}
\newcommand{\bG}{\bm{G}}
\newcommand{\bR}{\bm{R}}
\newcommand{\bF}{\bm{F}}
\newcommand{\bv}{\bm{v}}
\newcommand{\bu}{\bm{u}}

\newcommand{\bom}{\bm{\omega}}

\newcommand{\veps}{\varepsilon}
\newcommand{\be}{\begin{equation}}
\newcommand{\ee}{\end{equation}}

\newcommand{\transpose}{^{\operatorname{T}}}

\newcommand*\diff{\mathop{}\!\mathrm{d}}
\newcommand{\dt}{\partial_t}
\newcommand{\dx}{\nabla_{\bm{x}}}

\newcommand{\Gal}{\text{Gal}}

\begin{document}

\title{\Large{Integrating Machine Learning with Physics-Based Modeling}}

\author[1,2]{Weinan E}
\author[1]{Jiequn Han}
\author[2]{Linfeng Zhang} 
\affil[1]{Department of Mathematics, Princeton University}
\affil[2]{Program in Applied and Computational Mathematics, Princeton University}

\date{}


\maketitle
\begin{abstract}
Machine learning is poised as a very powerful tool that can drastically improve our ability to carry out
scientific research.
However, many issues need to be addressed before this becomes a reality.
This article
focuses on one particular issue of  broad interest: How can we integrate machine learning with
physics-based modeling to develop new interpretable and truly reliable physical models?
After introducing the general guidelines, we discuss the two most important issues for developing
machine learning-based physical models:  Imposing physical constraints and obtaining optimal datasets.
We also provide a simple and intuitive explanation for the fundamental reasons behind the success of modern machine learning,
as well as an introduction to the concurrent machine learning framework needed for integrating machine learning with physics-based modeling.
Molecular dynamics and moment closure of kinetic equations are used as examples to illustrate
the main issues discussed.
We end with a general discussion on where this integration will lead us to, and where the new frontier will be after machine learning is successfully integrated into scientific modeling.
\end{abstract}

\section{Fundamental laws and practical methodologies}

Physics is centered on two main themes:  the search for fundamental laws and the solution
of practical problems. The former has resulted in Newton's laws, Maxwell equations, the theory of relativity and quantum mechanics.
The latter has been the foundation of modern technology, ranging from automobiles, airplanes,
computers to cell phones. 
In 1929, after quantum mechanics was just discovered, Paul Dirac made the following claim \cite{dirac1929quantum}:

``The underlying physical laws necessary for the mathematical theory of a large part of physics and the whole of chemistry are thus completely known, and the difficulty is only that the exact application of these laws leads to equations much too complicated to be soluble. ''


What happened since then has very much confirmed Dirac's claim. It has been universally agreed that to understand problems
in chemistry, biology, material science, and engineering, one rarely needs to look further than quantum mechanics for first principles.
But solving practical problems using quantum mechanics principles, for example, the Schr\"odinger equation,
 is a highly non-trivial matter, due, among other things, to the many-body nature of the problem. 
To overcome these mathematical difficulties, 
researchers have proceeded along the following lines:
\begin{enumerate}
\item Looking for simplified models. For example, Euler's equations are often enough for studying gas dynamics.
There is no need to worry about the detailed electronic structure entailed in the Schr\"odinger equation.
\item  Finding approximate solutions using numerical algorithms and computers.
\item Multi-scale modeling:  In some cases one can model the behavior of a system at the macroscopic scale by using only a
 micro-scale model.
 \end{enumerate}

We will briefly discuss each.

\subsection{Seeking simplified models\label{sec:seeking}}
Seeking simplified models 
that either capture the essence of a problem or describe some phenomenon to a satisfactory accuracy
has been a constant theme in physics. 
Ideally, we would like our simplified models to have the following properties:
\begin{enumerate}
\item express fundamental physical principles (e.g. conservation laws),
\item obey physical constraints (e.g. symmetries, frame-indifference),
\item  be as universally accurate as possible:
Ideally one can conduct a small set of experiments in  idealized situations and obtain models that can be used
under much more general conditions,
\item be physically meaningful (interpretable).
\end{enumerate}
Euler's equations for gas dynamics is a very successful example of simplified physical models.
It is much simpler than
the first principles of quantum mechanics, and it is a very accurate model for dense gases.
For ideal gases,
the only parameter required is the gas constant. For complex gases, one needs  the entire equation
of state, which is a function of only two variables. 
Other success stories include the Navier-Stokes equations for viscous fluids, linear elasticity equations for
small deformations of solids, and the Landau theory of phase transition.

\begin{figure}[!ht]
  \centering
  \includegraphics[width=0.7\textwidth]{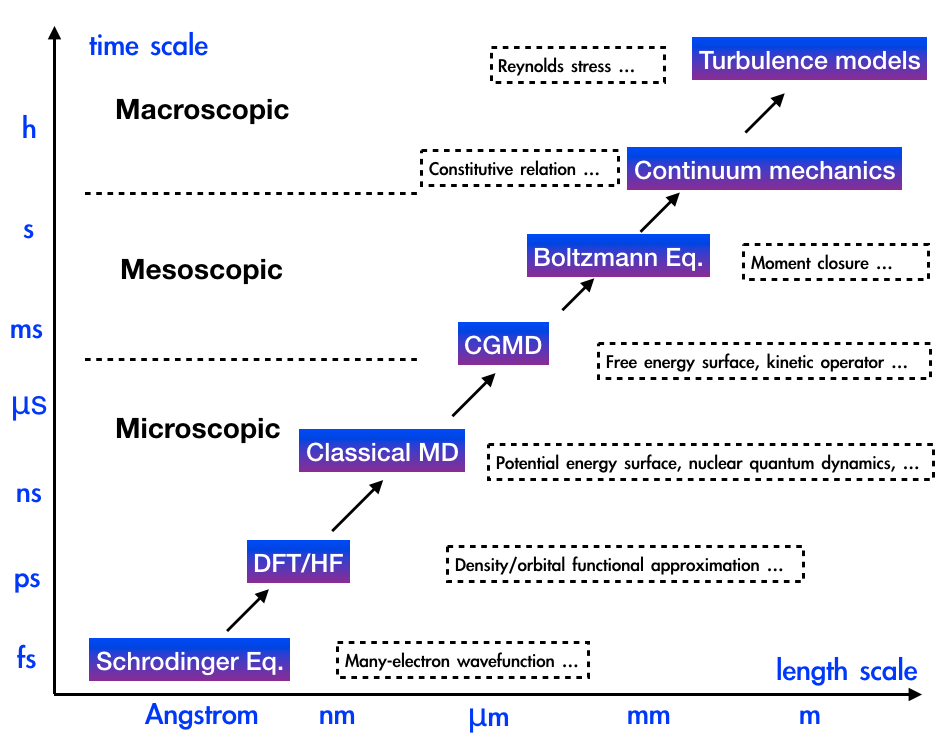}
\caption{Representative physical models at different scale and their most important modeling ingredients. \label{multi-scale-model}}
\end{figure}

Unfortunately, not every effort in coming up with simplified models has been as successful.
A good example is the effort of developing extended Euler equations for rarified gas. 
Since the work of Grad \cite{grad1949kinetic},
there have been  numerous efforts on developing 
Euler-like models for the dynamics of gases at larger Knudsen numbers.
So far this effort has not produced any widely accepted models yet.

To put things into perspective, we briefly discuss the methodologies that people use to obtain simplified models.

\begin{enumerate}
\item Generalized hydrodynamics \cite{degroot2013non}. 
The idea is to use symmetries, conservation laws, and the second law of thermodynamics to
extract as much definitive information about the dynamics as possible, and model the rest using linear constitutive relations.
A successful example is the Ericksen-Leslie equations for nematic liquid crystals, which works reasonably well
in the absence of line defects \cite{degennes1993physics}. 

\item The weakly nonlinear theory and gradient expansion trick.  This is an idea championed by Landau.  It has been
successfully used in developing models for phase transition, superconductivity, hydrodynamic instability, and a host
of other interesting physical problems.

\item Asymptotic analysis. This is a mathematical tool that can be used to systematically extract simplified models
by exploiting the existence of  some small parameters. For example, Euler's equation can be derived this way from the Boltzmann equation in the limit of small Knudsen number.

\item The Mori-Zwanzig formalism. This is a general strategy for eliminating unwanted degrees of freedom in a physical
model. The price to be paid is that the resulting model becomes nonlocal, for example, with memory effects.
For this reason, the application of the Mori-Zwanzig formalism has been limited to situations where the nonlocal
term is linear. An example is the generalized Langevin equations \cite{weinan2011principles}.

\item Principal component-based model reduction.  This is a broad class of techniques used in engineering for developing simplified models.
A typical example is to project the original model onto a few principal components obtained using data
produced from the original model. Well-known applications of this technique include reduced models that aim to capture
the dynamics of large scale coherent structures in turbulent flows \cite{tennekes1972first}.

\item  Drastically truncated numerical schemes. 
The well-known Lorenz system is an example of this kind. It
is the result of a drastic truncation of the two-dimensional incompressible 
Navier-Stokes-Boussinesq equation modeling thermal convection, an idealized model for weather prediction.
By keeping only three leading modes for the stream-function and temperature, one arrives at the system 
$$
\dot{x} = \sigma (y -x), \quad \dot{y} = -xz + r x -y, \quad \dot{z} = xy - bz,
$$
where $\sigma$ is the Prandtl number, $ r=Ra/Ra_c$ is the normalized Rayleigh number, and $b$ is some other parameter.
A well-known feature of this model is that it exhibits chaotic behavior, and this has been used as an indication
for the intrinsic difficulties in weather prediction. Of course, the obvious counter-argument is how well the Lorenz system
captures the dynamics of the original model.
\end{enumerate}


\subsection{Numerical algorithms}

Since analytical solutions are rare even after we simplify the models, one has to resort to numerical algorithms
to find accurate approximate solutions.  Many numerical algorithms have been developed for solving the partial
differential equations (PDEs) that arise from physics, including finite difference, finite element and spectral methods.
The availability of these algorithms has completely changed the way we do science,
and to an even greater extent, engineering.  For example nowadays numerical computation plays a dominant role
in the study of fluid and solid mechanics.  Similar claims can be made for atmospheric science, combustion, 
material sciences and a host of other disciplines, though possibly to a lesser extend.

Roughly speaking, one can say that we now have satisfactory algorithms for low dimensional problems (say three dimensions).
But things quickly become much more difficult as the dimensionality goes up.
A good example is the Boltzmann equation:  The dimensionality of the phase space and the nonlocality in 
the collision kernel makes it quite difficult to solve the Boltzmann equation using the kinds of algorithms mentioned above,
even though the dimensionality is small compared to the ones encountered in 
the many-body Schr\"odinger equation.

This brings us to the issue that lies at the core of the many difficult problems that we face: The
{\it curse of dimensionality (CoD)}:  As the dimensionality grows, the complexity (or 
computational cost) grows exponentially.

\subsection{Multi-scale modeling}

One important idea for overcoming the difficulties mentioned above is multi-scale modeling \cite{weinan2011principles},
a general philosophy based on modeling the behavior of macro-scale systems using reliable micro-scale models,
instead of relying on {\it ad hoc} macro-scale models.
The idea is to make use of the results of the micro-scale model on much smaller
spatial-temporal domains to predict the kind of macro-scale quantities that we are interested in \cite{weinan2011principles}. 
This general philosophy is valid for a wide range of scientific disciplines.
But until now, the success of multi-scale modeling has been less spectacular than what was expected twenty years ago.
The following challenges contributed to this.

\begin{enumerate}
\item The micro-scale models  are often not that reliable. For example when studying crack propagation, we often
use molecular dynamics as the micro-scale model. But the accuracy of molecular dynamics models for dynamic processes that involve bond breaking  is often questionable.


\item  Even though multi-scale modeling can drastically reduce the size of the micro-scale simulation required, it is still
beyond our current capability.

\item The key benefit we are exploiting with multi-scale modeling is the separation of the micro- and macro-scales 
of the problem.  But for the most interesting and most challenging problems, this often breaks down.

\item At a technical level, efficient multi-scale modeling requires effective algorithms for extracting the 
relevant information needed from micro-scale simulations.  This is a data analysis issue that has  not been adequately addressed.

\end{enumerate}

There are two basic strategies for multi-scale modeling, the sequential multi-scale modeling and the
concurrent multi-scale modeling.
In sequential multi-scale modeling, the needed components
 from the micro-scale model, for example, the constitutive relation,
 are obtained beforehand,
and this information is then supplied to some macro-scale model.  
For this reason, sequential multi-scale modeling
is also called ``precomputing''.
In concurrent multi-scale modeling, the coupling between the macro- and micro-scale models is done
on the fly as the simulation proceeds.
Sequential multi-scale modeling results in new models.
Concurrent multi-scale modeling results in new  algorithms.

\subsection{Difficulties that remain}
 A lot of progresses have been made using these methodologies in combination with physical insight as well as
 trial and error parameter fitting. This has enabled us to solve a wide variety of problems,
 ranging from performing density functional theory (DFT) calculations to predict properties of materials and molecules, to 
studying the climate using general circulation models. 
In spite of these advances, many issues remain as far as getting good models is concerned,
and many problems remain difficult.
Here are some examples:

\begin{enumerate}
\item A crucial component of the Kohn-Sham DFT is the exchange-correlation functional.
However, systematically developing efficient and accurate exchange-correlation functionals is still a very difficult task.

\item The most important component of a molecular dynamics model is the potential energy surface (PES) that
describes the interaction between the nuclei in the system. Accurate and efficient models of PES has always been
a difficult problem in molecular dynamics. We will return to this issue later.

\item An attractive idea for modeling the dynamics of macromolecules is to develop coarse-grained molecular dynamics models.
However, implementing this idea in practice is still a difficult task.

\item Developing hydrodynamic models for non-Newtonian fluids is such a difficult task that the subject itself has lost some steam.

\item Moment closure models for rarified gases. This was mentioned above and will be discussed in more detail later.

\item For fluids, we have the Navier-Stokes equations. What is the analog for solids? Besides linear elasticity models, there are hardly any other universally accepted continuum models of solids. This comment applies to nonlinear elasticity.  
Plasticity is even more problematic.

\item  Turbulence models.  This has been an issue ever since the work of Reynolds and we still do not have a systematic and
robust way of addressing this problem.
\end{enumerate}

From the viewpoint of developing models, one major difficulty has always  been the ``closure problem'':
When constructing simplified models, we encounter terms that need to be approximated in order
to obtain a closed system. Whether accurate closure can be achieved also depends in an essential way on
the level at which we impose the closure, i.e. the variables we  use to close the system.
For example, for DFT models, it is much easier to perform the required closure
for orbital-based models than  for orbital-free models. For turbulence models, the closure problem is much 
simplified in the context of large eddy simulation than Reynolds average equations, since a lot more information
is kept in the large eddy simulation.

From the viewpoint of numerical algorithms, these problems all share one important feature:  There are a lot of intrinsic degrees of freedom.  
For example turbulent flows are governed by the Navier-Stokes equations, which is a low
dimensional problem,  but its highly random nature means that we should really be looking for its statistical description,
which then becomes a very high dimensional problem with the dimensionality being proportional to the range of active scales.
Existing numerical algorithms can not handle these intrinsically high dimensional problems effectively.


In the absence of systematic approaches, one has to resort to {\it ad hoc} procedures which are not just
unpleasant but also unreliable. 
Turbulence modeling is a very good example of the kind of pain one has to endure in order to address  practical problems.

\subsection{Machine learning comes to the rescue}

Recent advance in machine learning offers us unprecedented power for approximating functions of many variables.
This allows us to go back to all the problems that were made difficult by  CoD.
It also provides an opportunity to reexamine the issues discussed above, with a new added dimension.

There are many different ways in which machine learning can be used to help solving problems that arise in
science and engineering.  We will focus our discussion on the following issue:
 \begin{itemize}
    \item How can we use machine learning to find new {\bf interpretable and truly reliable} physical models?
  \end{itemize}

While machine learning can be a very powerful tool, it usually does not work so well when used as  a blackbox.
One main objective of this article is to discuss the important issues that have to be addressed when integrating
machine learning with physics-based modeling.

Here is a list of basic requirements for constructing new physical models with the help of machine learning:
\begin{enumerate}
\item The models  should satisfy the requirements listed in Section~\ref{sec:seeking}.
\item The dataset that one uses to construct the model should be a good representation of all the practical situations that the model
is intended for.  It is one thing to fit some data. It is quite another thing to construct {\it reliable} physical models.
\item To reduce the amount of human intervention,  the process of constructing the models should be end-to-end. 
\end{enumerate}
Later we will examine these issues in more detail after a brief introduction to machine learning.


\section{Machine learning}

Before getting into the integration of machine learning with physics-based modeling, let us briefly discuss an issue
that is in many people's mind:
What is the magic behind neural network-based machine learning? 
While this question is still at the center of very intensive studies in theoretical machine learning, some insight can already
be gained from very simple considerations.


We will discuss a basic example, supervised learning:
Given ther dataset $ S=\{(\bx_j, y_j=f^*(\bx_j)), j=1, 2, \cdots, n \}$, learn $f^*$.
Here$\{\bx_j, j=1, 2, \cdots n\}$ is a set of points in the $d$ dimensional Euclidean space.  For simplicity we have 
neglected measurement noises.
 If the ``labels'' $\{y_j\}$ take discrete values, the problem is called a classification problem.
Otherwise it is called a regression problem.
In practice, one divides $S$ into two subsets, a training set used to train the model and a testing set
used to test the model.

A well-known example is the classification example of Cifar 10 dataset~\cite{krizhevsky2009learning}.
Here the task is to classify the images in the dataset into 10 categories. 
Each image has $32 \times 32$ pixels.
Thus it can be viewed as a point in $32 \times 32 \times 3 = 3072$ dimensional space. The last factor of $3$ comes from 
the dimensionality in the color space.  Therefore this problem can be viewed as approximating a discrete-valued function
of $3072$ variables.



The essence of the supervised learning problem is the approximation of a target function using a finite dataset. 
While the approximation of functions is an ancient topic, in classical approximation theory,  we typically use
 polynomials, piece-wise polynomials, wavelets, splines and other linear combinations of fixed basis functions
 to approximate a given function. These kinds of approximations suffer from CoD:
$$
f^* - f_m \sim m^{-\alpha/d} \Gamma(f^*),
$$
where $m$ is the number of free parameters in the approximation scheme,
$\alpha$ is some fixed constant depending on the approximation scheme, 
  $\Gamma(f^*)$ is an error constant depending on the target function $f^*$.
  Take $\alpha=1$.
  If we want to reduce the error by a factor of 10, the number of parameters needed goes up
  by $10^d$. This is clearly impractical when the dimensionality $d$ is large.

The one area in which we have lots of experiences and successes in handling high dimensional
problems is statistical physics, the computation of high dimensional expectations using Monte Carlo (MC) methods.
Indeed computing integrals of functions with millions of variables has become a routine practice in statistical 
physics that we forget to notice  how remarkable this is.
Let $g$ and $\mu$ be a function and a probability distribution in the $d$ dimensional Euclidean space respectively,
and let
$$I(g) = \E_{\bx \sim \mu}  g(\bx), \quad I_m(g) = \frac 1m  \sum_j g(\bx_j), $$
where $\{\bx_j\}$ is a sequence of i.i.d. samples of the probability distribution $\mu$. Then we have 
$$\EE(I(g)-I_m(g))^2 = \frac{\mbox{var}(g)}m, \quad 
\mbox{var}(g) = \int_X g^2(\bx) d\bx - \left(\int_X g(\bx) d\bx \right)^2.
$$
Note that the error rate $1/\sqrt{n}$ is independent of the dimensionality of the problem. Had we used grid-based algorithms
such as the Trapezoidal rule, we would have instead:
$$I(g) - I_m(g) \sim m^{-\alpha/d} \Gamma(g),$$
resulting in the CoD. 
Note also that  $\mbox{var}(g)$ can be very large in high dimensions. This is why
variance reduction is  a central theme in MC.

The MC story is the golden standard for dealing with high dimensional problems. 
A natural question is:  Can we
turn the function approximation problem, the key issue in supervised learning, into MC integration type of problems? 
To gain some insight, let us consider the Fourier representation of functions:
\begin{equation}
f^*(\bx) = \int_{\R^d} a(\bom) e^{i (\bom, \bx)} d \bom.
\end{equation}
We are used to approximate this expression by some grid-based discrete Fourier transform
\begin{equation}
f^*(\bx) \sim \frac 1m \sum_j a(\bom_j) e^{i (\bom_j, \bx)},
\end{equation}
and this approximation suffers from CoD.

If instead that we consider functions represented by 
\begin{equation}
f^*(\bx) = \int_{\R^d} a(\bom) e^{i (\bom, \bx)}  \pi(d\bom) =
\E_{\bom \sim \pi} a(\bom) e^{i (\bom, \bx)},
\label{MC-rep}
\end{equation}
where $\pi$ is a probability distribution on the Euclidean space $\R^d$, then the natural approximation becomes
\begin{equation}
f^*(\bx) \sim  \frac 1m \sum_{j=1}^m a(\bom_j) e^{i (\bom_j, \bx)},
\label{MC-rep-discrete}
\end{equation}
where $\{\bom_j\}$ are  i.i.d. samples of $\pi$. For reasons discussed earlier, this approximation does not suffer from CoD.
The right hand side of \eqref{MC-rep-discrete} is an example of neural network functions with one hidden layer
and activation function $\sigma$ defined by $\sigma(z) = e^{iz}$.
If one wants a one-sentence description of the magic behind neural network models, the reasoning above
should be a good candidate.

There is one key difference between the supervised learning problem and MC integration:
In MC, the probability distribution $\mu$ is given, usually in fairly explicit form.
In supervised learning, the probability distribution $\pi$ is unknown. Instead we are only given information
about the target function $f^*$ on a finite dataset. 
Therefore it is infeasible to perform the approximation in \eqref{MC-rep-discrete}. Instead one has to replace
it by some optimization algorithms based on fitting the dataset.


\section{Concurrent machine learning}

In classical supervised learning,  the labeled dataset is given beforehand and learning is performed afterwards.
This is usually not the case when machine learning is used in connection with physical models.
Instead, data  generation and training is an interactive process: Data is generated and labeled on the fly as 
model training proceeds.  In analogy with multi-scale modeling, we refer to the former class of problems
``sequential machine learning'' problems and the latter kind ``concurrent machine learning'' problems.

Active learning is something in between.  In active learning, we are given the unlabeled data, and we decide
which ones should be labeled during the training process. In concurrent learning, one also needs to generate the
unlabeled data.

To come up with reliable physical models, one has to have reliable data:  The dataset should be able to 
represent all the situations that the model is intended for. On the other hand, 
since data generation typically involves
numerical solutions of the underlying micro-scale model, a process that is often quite expensive, we would like
to have as small a dataset as possible.
This calls for a procedure that generates the data adaptively in an efficient way.

\subsection{The ELT algorithm}

The (exploration-labeling-training) ELT algorithm was first formulated in Ref.~\cite{zhang2018reinforced}
although similar ideas can be traced back for a long time.
Starting with no (macro-scale) model and no data but with a micro-scale model, the ELT algorithm
proceeds iteratively with the following steps:
\begin{enumerate}
\item {\bf exploration}: explore the configuration space, and decide which configurations need to be labeled.
The actual implementation of this step requires an algorithm for exploring the configuration space
and a criterion for deciding which configurations need to be labeled.

Oftentimes the current macro-scale model is used to help with the exploration.

\item {\bf labeling}: compute the micro-scale solutions for the configurations that need to be labeled, and place them in the training dataset.

\item {\bf training}: train the needed macro-scale model. 
\end{enumerate}

The ELT algorithm requires the following components:  a macro-scale explorer, an indicator to evaluate whether
a given configuration should be labeled, a micro-scale model used for the labeling, and a machine learning model
for the quantities of interest.

One way of deciding whether a given configuration needs to be labeled is to estimate the error of the
current machine learning model at the given configuration. When neural network models are used, 
a convenient error indicator can be constructed as the variance between the predictions given by an ensemble of neural network models (say with the same architecture but different initialization of the optimization algorithm).
If the variance is small, then the predictions from different models are close to each other.
In this case, it is likely that these predictions are already quite accurate at the given configuration.  
Therefore this configuration does not need to be labeled.  
Otherwise, it is likely that the predictions are inaccurate and the given configuration should be labeled.

\subsection{Variational Monte Carlo}
Another example where concurrent learning might be relevant is machine learning-based approach for variational 
Monte Carlo algorithms.
Variational Monte Carlo (VMC) \cite{mcmillan1965VMC,ceperley1977VMC}, for solving the Schr\"odinger equation  was among the first set of applications
of machine learning in computational science~  \cite{carleo2017solving,weinan2018deepritz}.  
In these and other similar applications, one
begins with no data and no solution, the data is generated on the fly as the solution process proceeds.
Given a Hamiltonian $\hat{H}$ with the state variable $\bx$, the variational principle states that the wave-function $\Psi$ associated with the ground state minimizes within the required symmetry the following energy functional 
\begin{align}
E[\Psi] =
\frac{\int\Psi^{*}(\bx)\hat{H}\Psi(\bx)\,d\bx}{\int\Psi^{*}(\bx)\Psi(\bx)\,d\bx}
=\E_{\bx\sim |\Psi(x)|^2}
\frac{\hat{H}\Psi(\bx)}{\Psi(\bx)}.
\label{eq:vmc}
\end{align}
In a machine learning-based approach, one parameterizes 
 the wave-function using some machine learning model, $\Psi \sim \Psi_{\theta}$ with some parameter $\theta$
 that needs to determined.
In this case  both the integrand and the distribution in \eqref{eq:vmc} depend on $\theta$. 
To find the optimal $\theta$, the learning process consists of two alternating phases, 
 the sampling phase and the optimization phase.
In the sampling phase, a sampling algorithm is used to sample the probability distribution generated by the current 
approximation of  $\Psi^2$.  This sample is used to evaluate the loss function in \eqref{eq:vmc}.
In the optimization phase, one iteration of the optimization algorithm is applied to the loss function.
This procedure also resembles the EM (expectation-maximization) algorithm.

\section{Molecular modeling }

One of the most successful applications of machine learning to scientific modeling is in molecular dynamics.
Molecular dynamics (MD) is a way of studying the atomic scale property of materials and molecules by tracking the dynamics of the nuclei in the system using classical Newtonian dynamics.  
The key issue in MD is how to model the potential energy surface (PES) that describes the interaction between the nuclei.  
This is a function that depends on the position of all the nuclei in the system:
\[
E=E(\bm{x}_{1},\bm{x}_{2},...,\bm{x}_{i},...,\bm{x}_{N}),
\]
Traditionally, there have been two ways to deal with this problem. 
The first is to compute the inter-atomic forces on the fly using first principles-based models such as the DFT~\cite{kohn1965self}.
This is the so-called {\it ab initio} molecular dynamics pioneered by  Car and Parrinello \cite{car1985unified}.
This  approach gives an accurate description of the system under consideration.
However, it is computationally very expensive, limiting the size of the system that one can handle to thousands of atoms.
At the other extreme is the approach of using empirical potentials. 
In a nutshell this approach aims at modeling the PES with empirical formulas.  
A well-known example is the Lennard-Jones potential, a reasonable model for describing the interaction between inert atoms:
\[
V_{ij}=4\epsilon \left((\frac{\sigma}{r_{ij}})^{12}-(\frac{\sigma}{r_{ij}})^6\right), \quad E=\frac{1}{2}\sum_{i\not=j}V_{ij}.
\]
Empirical potential-based MD is very efficient, but guessing the right formula that can model the PES accurately enough is understandably a very difficult task, particularly for complicated systems such as high entropy alloys.
In practice one has to resort to a combination of physical insight and {\it ad hoc} approximation.
Consequently the gain in efficiency is at the price of reliability. 
In particular, since these empirical potentials are typically calibrated using  data for equilibrium systems, their accuracy at other interesting configurations, such as transition states, is often in doubt.  

The pioneering work of Behler and Parrinello introduced a new paradigm for performing {\it ab initio} molecular dynamics \cite{behler2007generalized}. 
In this new paradigm,  a first principle-based model acts as the vehicle for generating highly accurate data. 
Machine learning is then used to parametrize this data and output a model of the PES with {\it ab initio} accuracy.  
See also the work of Csanyi et al \cite{bartok2010gaussian}.

\begin{figure}[!ht]
  \centering
  \includegraphics[width=0.6\textwidth]{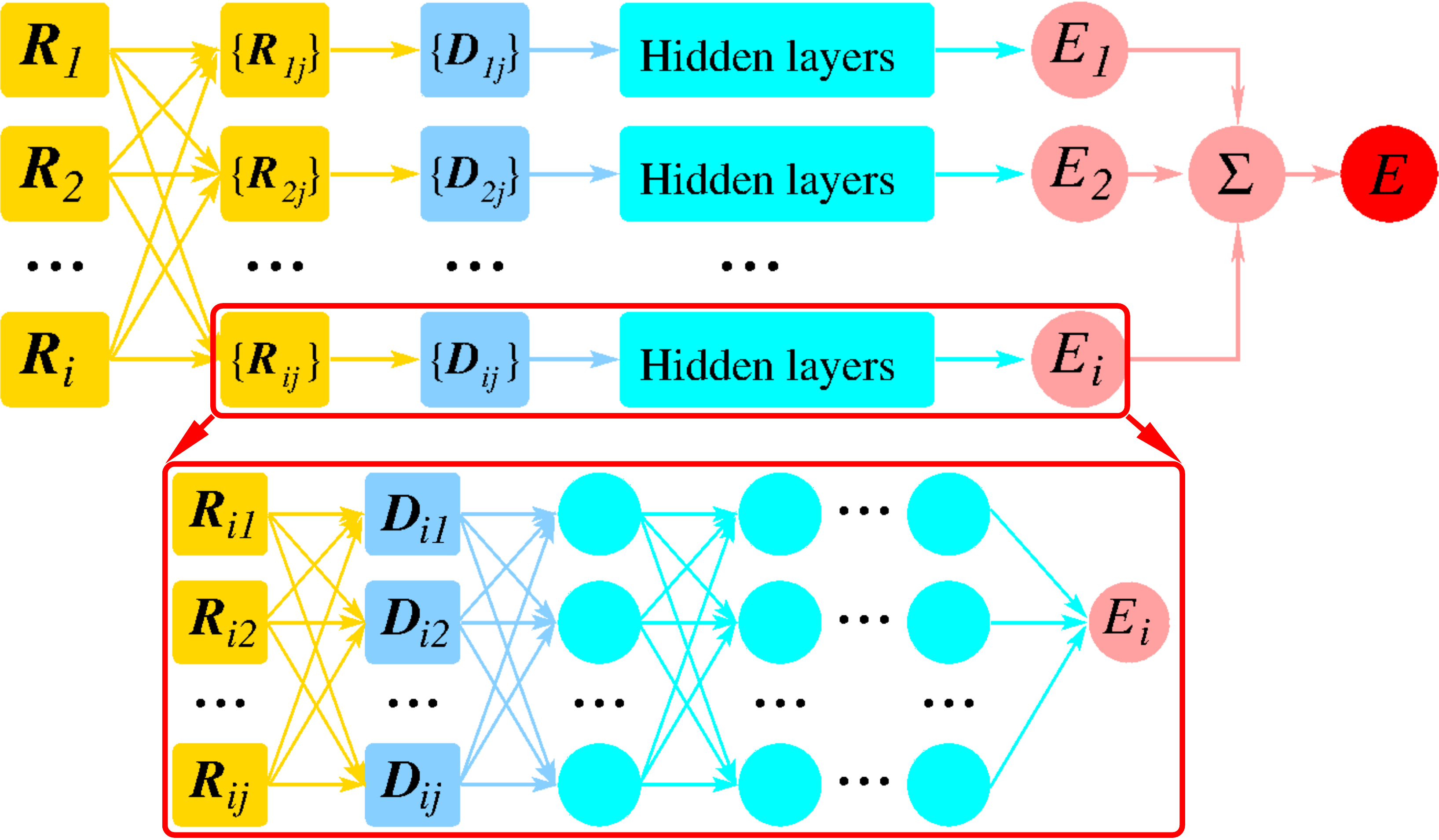}
\caption{
Illustration of the neural network architecture for a system of $N$ atoms.
From Ref.~\cite{zhang2018deep}.
\label{composite}}
\end{figure}

Behler and Parrinello introduced a neural network architecture that is naturally extensive (see Figure~\ref{composite}).
In this architecture, each nucleus is associated with a subnetwork. 
When the system size is increased, one just has to append the existing network with more subnetworks corresponding to the new atoms added.

As discussed before, to construct truly reliable models of PES, one has to address two problems. The first is getting good data.  The second is imposing physical constraints.
   
Let us discuss the second problem first. 
The main physical constraints for modeling the PES are the symmetry constraints.
Naturally the PES model should be invariant under translational and rotational symmetries.
It should also be invariant under the relabeling of atoms of the same chemical species. 
This is a permutational symmetry condition. 
Translational symmetry is automatically taken care of by putting the origin of the local coordinate frame for each nucleus at that nucleus.  
To deal with other symmetry constraints, Behler and Parrinello resorted to constructing ``local symmetry functions'' using bonds, bond angles, etc., in the same spirit as the constructions in the Stillinger-Weber potential~\cite{stillinger1985SW}. 
This construction is a bit {\it ad hoc} and is vulnerable to the criticism mentioned earlier for empirical potentials.
   
\begin{figure}
\centering
 \includegraphics[width=0.96\textwidth]{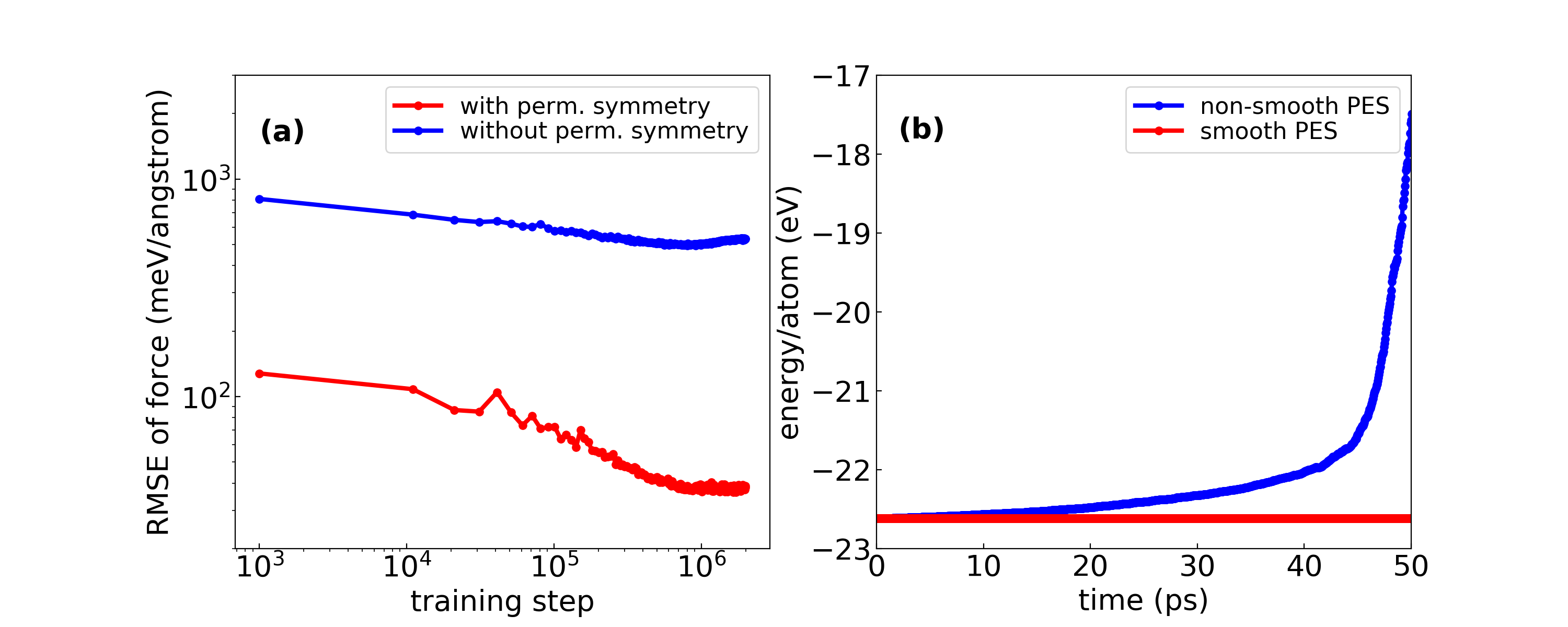}
 \caption{(a) Test accuracy without imposing the permutational symmetry (blue) and with a poor man's way of enforcing the permutational constraints (red).
 (b) Total energy per atom as a function of time from microcanonical DeePMD simulations, using a poor man's way to impose symmetry constraints (blue) and using the smooth embedding network (red).
\label{symm-1}}
\end{figure}

To demonstrate the importance of preserving the symmetry,  we show in Figure~\ref{symm-1} the
comparison of the results with and without imposing symmetry constraints.  
One can see that without enforcing the symmetry, the test accuracy of the neural network model is rather poor.
 Even a poor man's version of enforcing symmetry can improve the test accuracy drastically.

A poor man's way of enforcing symmetry is to remove by hand the degrees of freedom associated with these symmetries.  
This can be accomplished as follows~\cite{han2017deep,zhang2018deep}:
\begin{itemize}
\item enforcing  rotational symmetry by fixing in some way a local frame of reference.
For example, for atom $i$, we can use its nearest neighbor to define the $x$-axis and use the plane spanned by its two nearest neighbors to define the $z$-axis, thereby fixing the local frame.
\item enforcing  permutational symmetry by fixing an ordering of the atoms in the neighborhood.
For example, 
within each species we can sort the atoms according to their distances to the reference atom $i$.
\end{itemize}
This simple operation allows us to obtain neural network models with very good accuracy, as shown in Figure \ref{symm-1} (a).

The only problem with this procedure is that it creates small discontinuities when the ordering of the atoms in a neighborhood changes. 
These small discontinuities, although negligible for sampling a canonical ensemble, show up drastically in a microcanonical molecular dynamics simulation, as can be seen in Figure \ref{symm-1} (b).

To construct a smooth potential energy model that satisfies all the symmetry constraints 
 we have to reconsider how to represent the general form of symmetry-preserving  functions.
The idea is to precede the fitting network in each subnetwork by an embedding network that produces 
a sufficient number of symmetry-preserving functions.  In this way the end model is automatically symmetry-preserving.
The PES generated this way is called (the smooth version of) Deep Potential \cite{zhang2018end}.
MD driven by Deep Potential is called DeePMD.
The idea of embedding network is fairly general and can be used in other situations where symmetry is an important issue~\cite{zhang2020dw,grace2020raman}.


To construct the general form of symmetry-preserving functions, we draw inspiration from the following two observations.

{\bf Translation and Rotation.}  The matrix
$
\Omega_{ij}\equiv \bm{r}_i\cdot\bm{r}_j
$
is an over-complete array of invariants with respect to translation and rotation \cite{bartok2013representing,weyl2016classical}, 
i.e., it contains the complete information of the point pattern $\bm{r}$.
However, this symmetric matrix switches rows and columns under a permutational operation. 

{\bf Permutation.} We consider a generalization of Theorem 2 of Ref.~\cite{zaheer2017deep}: A function $f(\bm{r}_{1},...,\bm{r}_{i},...,\bm{r}_{N})$ is invariant to the permutation of instances in $\bm{r}_{i}$, if and only if it can be decomposed into the form $\rho(\sum_{i}g(\bm{r}_{i})\bm{r}_{i})$, for suitable transformations $g$ and $\rho$.

\begin{figure}[!ht]
\centering
 \includegraphics[width=0.5\textwidth]{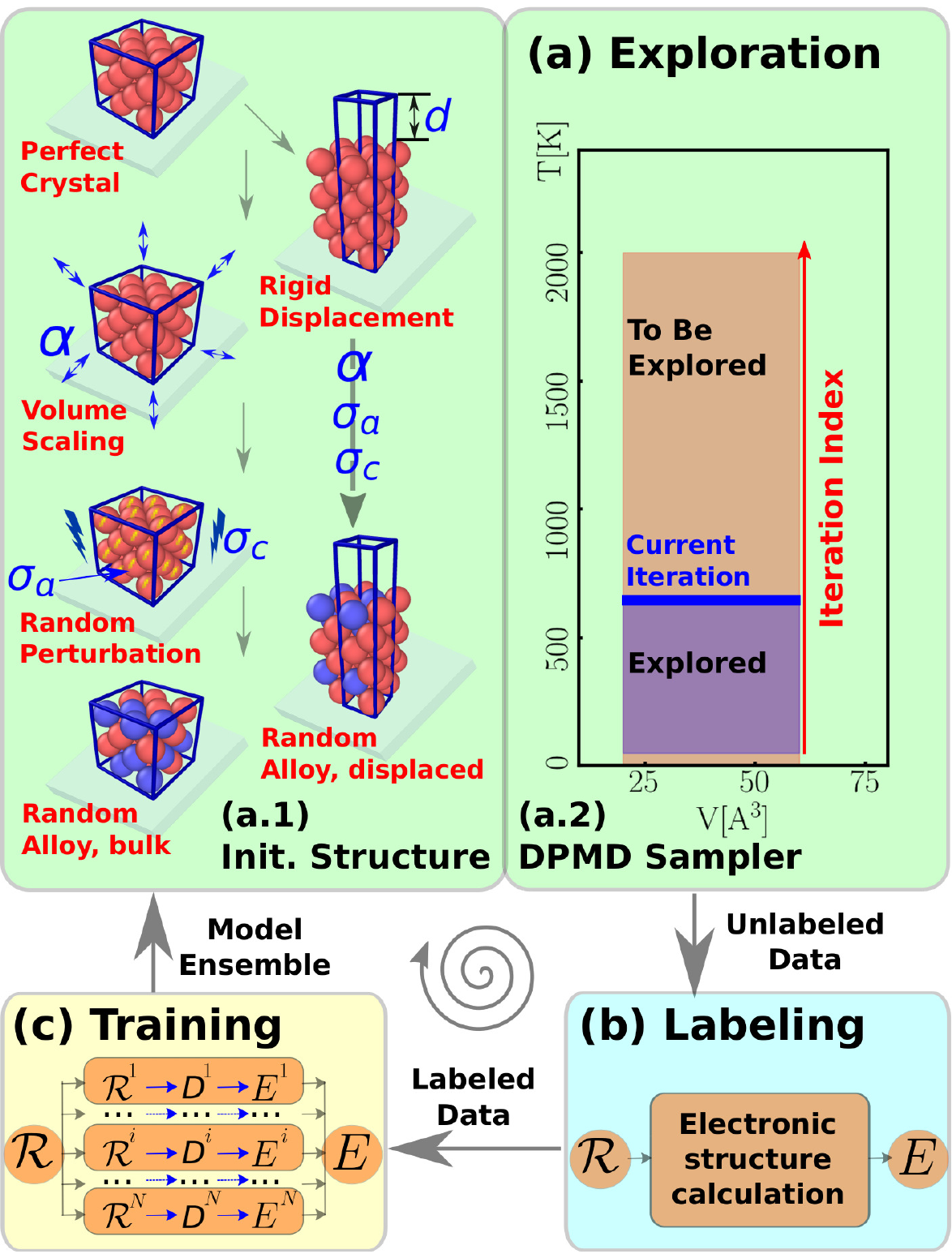}
\caption{
Schematic plot of one iteration of the DP-GEN scheme, taking the Al-Mg system as an example.
(a) Exploration with DeePMD.
(a.1) Preparation of initial structures. We start from stable crystalline structures of pure Al and Mg, compress and dilate the stable structures uniformly to allow for a larger range of number densities, and then randomly perturb the atomic positions and cell vectors of all the initial crystalline structures.
Surface-related structures are generated with rigid displacement.
Based on configurations of pure metal we also generate random alloy structures.
(a.2) Canonical simulation at a given temperature. 
(b) Labeling with electronic structure calculations.
(c) Training with the DP model.
From ~\cite{zhang2019active}.
\label{DP-GEN}}
\end{figure}

We now go back to the first problem, the generation of good data using the ELT algorithm.
The objective is to develop a procedure that can generate Deep Potentials for a given system that can be used in a wide variety of thermodynamic conditions.
In the implementation in~\cite{zhang2019active},  exploration was done using the following ideas. 
At the macro-scale level, one samples the $(T, p)$ space of thermodynamic variables in some way.
For each value of $(T, p)$, one samples the canonical ensemble, using MD with the Deep Potential available at the current  iteration. 
In addition, the efficiency of the exploration can be improved by  initializing the MD using a variety of different initial configurations, say different crystal structures.
Labeling was done using the Kohn-Sham DFT with periodic boundary condition.
Training was done using Deep Potential.
An example is given in Figure~\ref{DP-GEN}.  
In this example, only $\sim$0.005\% of the configurations explored was labeled~\cite{zhang2019active}.
In particular, the Deep Potential constructed this way satisfies all the requirements listed in Section 1.5.

\begin{figure}[!ht]
\centering
 \includegraphics[width=0.8\textwidth]{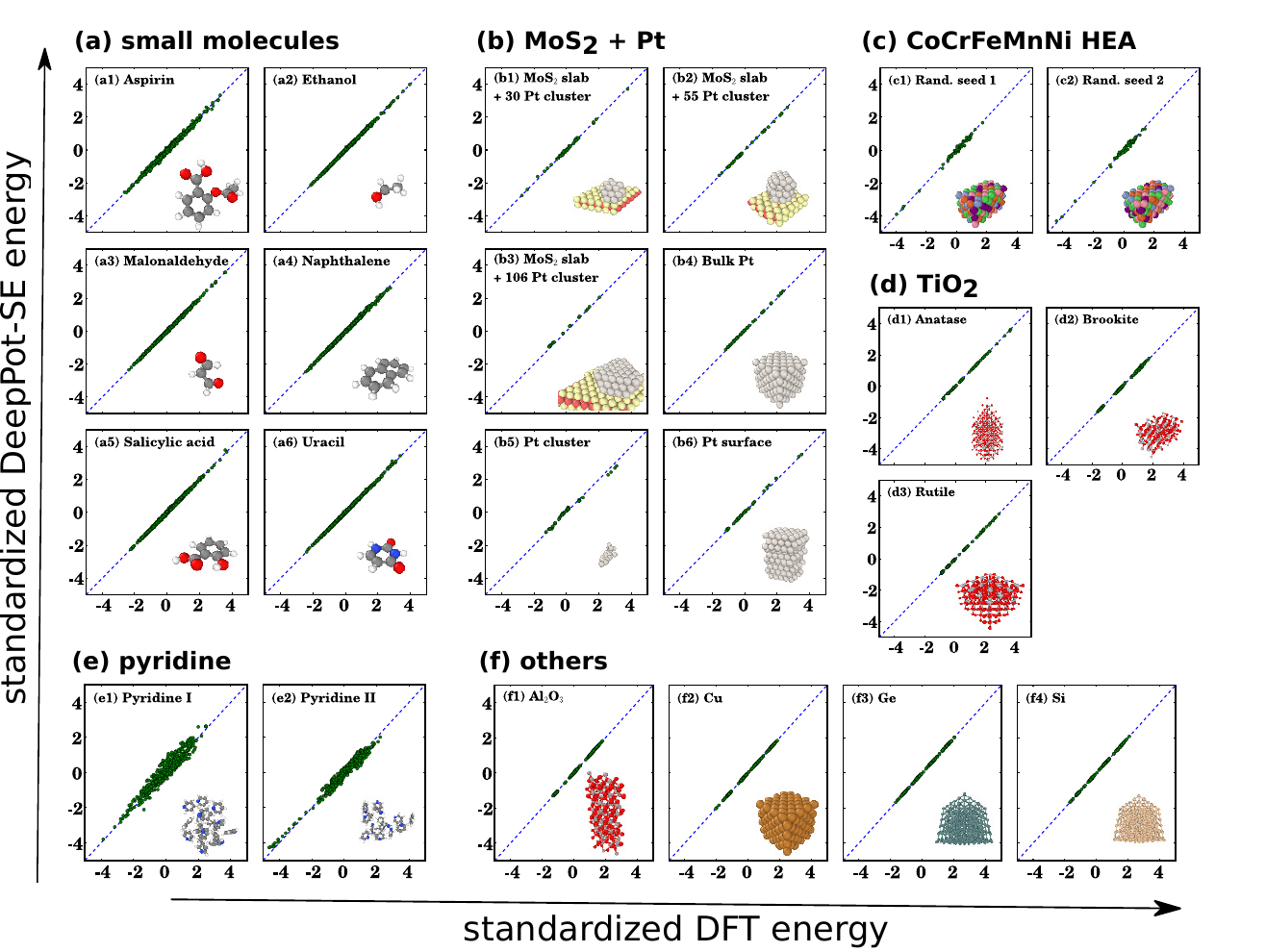}
\caption{
Comparison of the DFT energies and the Deep Potential-predicted energies on the testing snapshots. 
The energies are standardized for a clear comparison.
(a) Small molecules. 
(b) MoS${}_2$ and Pt. 
(c) CoCrFeMnNi high-entropy alloy (HEA). 
(d) TiO${}_2$.
(e) Pyridine (C${}_5$H${}_5$N).
(f) Other systems: Al${}_2$O${}_3$, Cu, Ge, and Si. 
From~\cite{zhang2018end}.
\label{DP-1}}
\end{figure}

\begin{figure}[!ht]
\centering
 \includegraphics[width=0.8\textwidth]{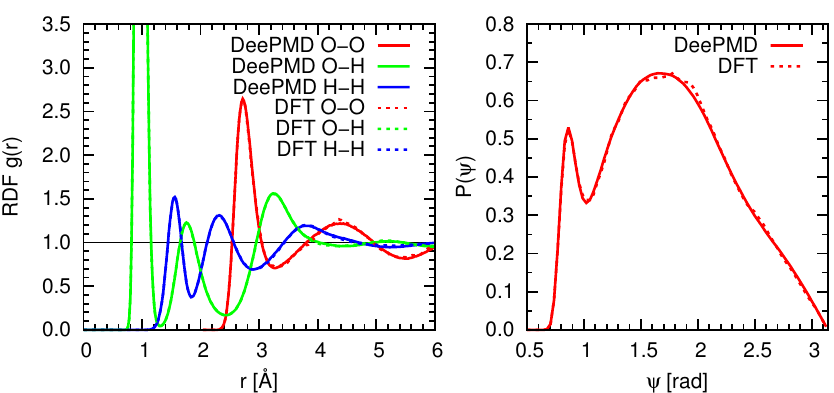}
\caption{Correlation functions of liquid water from DeePMD and PI-AIMD. Left: radial distribution functions. Right: the O-O-O angular distribution function.
From ~\cite{zhang2018deep}.
\label{DP-2}}
\end{figure}

As a first example of the applications of these ideas, we show in Figure \ref{DP-1} the accuracy of the Deep Potential for a wide variety of systems, from small molecules, to large molecules, to oxides and high entropy alloys.  One can see that in all the cases studied, Deep Potential
achieves an accuracy comparable to that of the underlying DFT.
As a second example, we show the results of using Deep Potential  to study the structural information of  liquid water (see Figure~\ref{DP-2}).
One can see  that the results  for the radial and angular distribution functions are comparable to the results from {\it ab initio} MD.
As a third example, combined with the state-of-art high performance computing platform (Summit), molecular dynamics simulations with {\it ab initio} accuracy have been pushed to systems of up to 100 million atoms, making it possible to study more complex phenomena that require truly large-scale simulations~\cite{lu202086,jia2020pushing}.
An application to nanocrystalline copper was presented in Ref.~\cite{jia2020pushing}.
Deep Potential and DP-GEN have been implemented in the open-source packages DeePMD-kit~\cite{wang2018kit} and DP-GEN~\cite{zhang2020dpgen}, respectively, and have attracted researchers from various disciplines.

For other related work, we refer to~\cite{podryabinkin2017active,smith2018less,bernstein2019novo}.

\section{Moment closure for kinetic models of gas dynamics}


  

The dynamics of gases is described very well by the Boltzmann equation
for the single-particle phase space density function $f=f(\bx, \bv, t): \R^3\times \R^3\times \R \rightarrow \R^{+}$:
\begin{equation}
\label{eq:Boltzman}
  \partial_t f + \bv\cdot \nabla_{\bx} f = \frac{1}{\veps} Q(f).
\end{equation}
Here we use $\varepsilon $  to denote  the Knudsen number, $Q$ is the collision operator.
When $\varepsilon$ is small, the Boltzmann equation can be accurately reduced to the Euler equation for the
macroscopic variables, $\rho$, $\bu$ and $T$  which represent density, bulk velocity and temperature respectively:
\begin{equation}
  \rho = \int_{\R^3} f\diff\bv, \quad \bu = \frac{1}{\rho}\int_{\R^3} f\bv\diff\bv, \quad T = \frac{1}{3\rho}\int_{\R^3} f |\bv - \bu|^2\diff\bv.
  \label{moments-1}
\end{equation}
\begin{equation}
  \partial_t \bU + \nabla_{\bx} \cdot \bF(\bU) = 0,
\end{equation}
with $\bU = (\rho, \rho\bu, E)^T $, $\bF(\bU) = (\rho\bu, \rho \bu\otimes\bu + pI, (E + p)\bu)^T $, 
$p = \rho T$, $E = \frac{1}{2}\rho |\bu|^2 + \frac{3}{2}\rho T$.
Euler's equation can be obtained by projecting Boltzmann's equation onto the first few moments that defined in 
\eqref{moments-1}, and making use of the local Maxwellian approximation to close the system:
\begin{equation}
\label{eq:Maxwellian}
  f_M(\bv) = \frac{\rho}{(2\pi T)^{\frac{3}{2}}}\exp\left(-\frac{|\bv - \bu|^2}{2T}\right).
\end{equation}

The accuracy of the Euler's equation deteriorates when the Knudsen number increases, since the local Maxwellian
is no longer a good approximation for the solution of the Boltzmann equation.
To improve upon Euler's equation, Grad proposed the idea of using more moments to arrive at extended Euler-like
equations \cite{grad1949kinetic}. As an example,  Grad proposed the well-known 13-moment equation. Unfortunately Grad's 13-moment
equation suffers from all kinds of problems, among which is the loss of hyperbolicity in certain regions of the
state space. Since then, many different proposals have been made in order to obtain reliable moment closure systems.
The usual procedure is as follows.
\begin{enumerate}
\item Start with a choice of a finite-dimensional linear subspace of functions of $\bv$ (usually some set of polynomials, e.g., Hermite polynomials).

\item Expand $f(\bx, \bv, t)$  using these functions as bases and take the coefficients as moments (including macroscopic variables $\rho$, $\bu$, $T$, etc.).

\item Close the system using some simplified assumptions, e.g., truncating moments of higher orders.
\end{enumerate}

For instance, in Grad 13-moment system, the moments are constructed using the basis $\{1, \bv, (\bv - \bu)\otimes(\bv - \bu), |\bv - \bu|^2(\bv - \bu)\}$.
It is fair to say that at the moment, this effort has not really succeeded.

There are two important issues that one has to address in order to obtain reliable moment closure systems.
The first is: What is the best set of moments that one should use? The second is: How does one close the projected system?
The second problem is the well-known closure problem. 


One possible machine learning-based approach is as follows \cite{han2019uniformly}.


1.  Learn the set of generalized moments using some machine learning models such as the auto-encoder.
The problem can be formulated as follows:
Find an encoder $\Psi$ that maps $f(\cdot, \bv)$ to generalized moments $\bW \in \R^M$ and a decoder $\Phi$ that recovers the original $f$ from $\bU, \bW$
$$\bW=\Psi(f)=\int \bw f\diff \bv, \quad \Phi(\bU, \bW)(\bv) = h(\bv; \bU, \bW).$$

The goal is to find an optimal $\bw$ and $h$ parametrized by neural networks through minimizing
$$\underset{f\sim \mathcal{D}}{\E} \|f - \Phi(\Psi(f))\|^2
    + \lambda_\eta (\eta(f) - h_\eta(\bU, \bW))^2.$$
where $\eta(f)$ denotes the entropy. 

We call $(\bU, \bW)$ the set of generalized moments.

{2.  Learn the  reduced model for the set of generalized moments.}

Projecting the kinetic equation onto the set of generalized moments, one obtains equations of the form:
\begin{equation}
  \left\{
  \begin{aligned}
     & \dt \bU + \dx\cdot \bF(\bU, \bW; \veps) = 0, \\
     & \dt \bW + \dx\cdot \bG(\bU, \bW; \veps) = \frac{1}{\veps}\bR(\bU, \bW; \veps).
  \end{aligned}
  \right.
\end{equation}
This is the general conservative form of the moment system. The term $1/\veps$ is inherited directly from \eqref{eq:Boltzman}. 
Our task is to learn $\bF, \bG, \bR$ from the original kinetic equation.


To get reliable reduced models, we still have to address the issues of getting good data and respecting physical 
constraints.
To address the first problem, one can use the ELT algorithm with the following implementation \cite{han2019uniformly}:
\begin{itemize}
\item exploration:  random initial conditions made up of random waves and discontinuities.
\item labeling: solving the kinetic equation.  This can be done locally in the momentum space since the kinetic models
have a finite speed of propagation.
\item training:  as discussed above.
\end{itemize}

To address the second problem,  one notices that there is an additional symmetry: the Galilean invariance. This is a dynamic constraint respected by the kinetic equation. Specifically, for every $\bu\in \R^3$, define
$$f'(\bx, \bu, t) = f(\bx-t\bu', \bv-\bu', t).$$
If $f$ is a solution of the Boltzmann equation, then so is $f'$.
To enforce similar constraints for the reduced model, we define the Galilean-invariant moments by:
\begin{equation}
    \bW_\Gal=\Psi(f)=\int f(\bv)\bw\left(\frac{\bv-\bu}{\sqrt{T}}\right)\diff \bv.
\end{equation}
Modeling the dynamics of $\bW_\Gal$ becomes more subtle due to the spatial dependence in $\bu, T$.
For simplicity we will work with a discretized form of the dynamic model in the closure step.
Suppose we want to model the dynamics of $\bW_{\Gal,j}$ at the spatial grid point indexed by $j$. Integrating the Boltzmann equation 
against  the generalized basis at this grid point gives
{\small{
\begin{equation}
    \partial_t \int_{\R^D}f(\bv)\bw\left(\frac{\bv-\bu_j}{\sqrt{T_j}}\right)\diff\bv + \dx \cdot \int_{\R^D}f(\bv)\bw\left(\frac{\bv-\bu_j}{\sqrt{T_j}}\right)\bv\transpose \diff \bv = \int_{\R^D}\frac1\veps Q(f)\bw\left(\frac{\bv-\bu_j}{\sqrt{T_j}}\right)\diff \bv.
\end{equation}
}}
The closes equations now take the form
\begin{equation}
    \partial_t \bW_\Gal + \dx \cdot \bG_\Gal(\bU, \bW_\Gal; \bU_j) = \frac{1}{\veps}\bR_\Gal(\bU, \bW_\Gal).
\end{equation}
One can perform machine learning using this form.



\begin{figure}[!ht]
\centering
 \includegraphics[width=0.99\textwidth]{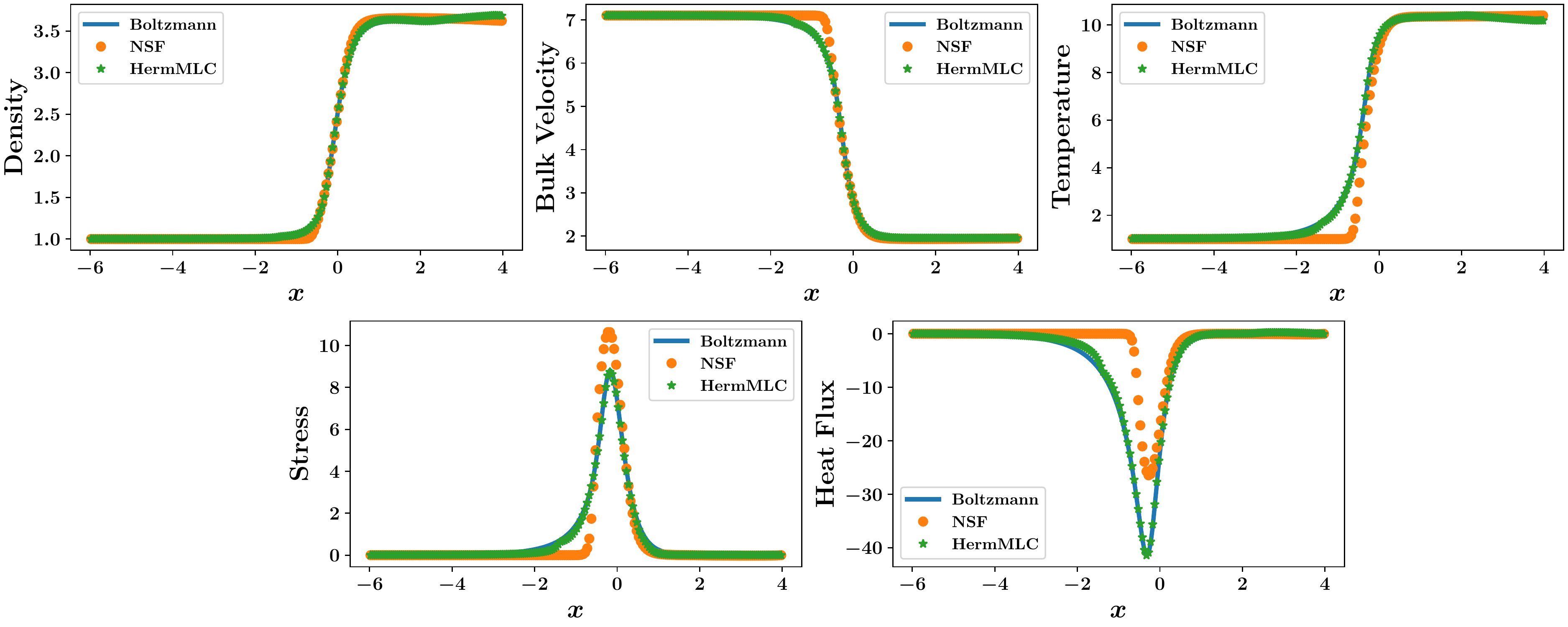}
\caption{Profiles of mass density, bulk velocity, temperature, normal stress, and heat flux when the Mach number equals 5.5, obtained from the Boltzmann
equation, the Navier-Stokes-Fourier equations, and HermMLC.
\label{boltz_shock}}
\end{figure}

The numerical results given in \cite{han2019uniformly} shows the promise of this approach. 
Here we only give an example of using this approach  to study shock wave structure in
 high-speed rarefied gas flows, when the gas experiences a fast transition in a tube between two equilibrium states.
During the transition the flow deviates substantially  from the thermodynamical equilibrium.
The classical  Navier-Stokes-Fourier (NSF) equations  fail  when the Mach number is bigger than 2.
Figure~\ref{boltz_shock} presents a comparison of the shock wave profiles with the BGK (Bhatnagar–Gross–Krook) collision
model~\cite{bhatnagar1954model} when the Mach number is 5.5,
and the results of the machine learning-based closure model with 3 additional Hermite moments (HermMLC).
One can see that the machine learning-based model achieves good agreements with the original Boltzmann equation.

\section{Discussions and concluding remarks}
We have focused on developing sequential multi-scale models using current machine learning.
These models are very much like the physical models we are used to,
except that some functions in the models exist in the form of machine learning models such as neural  network
functions.  
In this spirit, these models are no different from Euler's equations for complex gases where
the equation of state is stored as tables or subroutines.
As we discussed above, following the right protocols will ensure that these machine learning-based models are just as
reliable as the original micro-scale models.

In principle the same procedure and same set of protocols are also applicable in a purely data-driven context without
a micro-scale model: One just has to replace the labeler by experimental results.
In practice, this is still a relatively unexplored territory.

Another way to integrate machine learning with physics-based modeling is from the viewpoint of learning dynamical systems.
After all,  the physical models we are looking for are all examples of dynamical systems.
One representative machine learning model for learning dynamical systems is the recurrent neural networks \cite{rumelhart1986learning}.
In this connection, an 
 interesting observation made in \cite{ma2018model} is that there is a natural connection between the Mori-Zwanzig formalism
and the recurrent neural network model. Recurrent neural network models are machine learning models for time series.
It models the output time series using a local model by introducing hidden variables. 
In the absence of these hidden variables, the relationship between the input and output time series will be nonlocal 
with memory effects.  In the presence of these hidden variables (which are learned as part of the training process),
the relationship becomes local. 
Therefore, recurrent neural network models can be viewed as a way of unwrapping the memory effects by introducing hidden variables.
One can also think of it as a way of performing Mori-Zwanzig by compressing the space of unwanted degrees of freedom.
This connection and this viewpoint of integrating machine learning with physics-based modeling has yet to be explored.

Going back to the seven problems listed in Section 1.4, we have already addressed problems 2 and 5.
For problem 1, some progress has been reported in \cite{chen2020deephf} and \cite{cheng2019universal,nagai2020completing} for molecules.
Similar models/algorithms as DeePMD and Deep Potential have been developed for coarse-grained molecular
dynamics \cite{zhang2018deepcg}.
For problem 4,  some initial progress has been made in \cite{lei2020machine}.
For problem 7, efforts that use neural network models can be traced back to \cite{sarghini2003neural} but there is still a huge gap
compared with the standards promoted in this paper.  
Problem 6 is wide open.
Overall, one should expect substantial progress to be made on all these problems in the next few years.

This brings us to a new question: What will be the next frontier, after machine learning is successfully integrated
into physics-based models?  With the new paradigm firmly in place, the key bottleneck will quickly become
the micro-scale models that act as the generators of golden standard data.  In the area of molecular modeling,
this will be the solution of the original many-body Schr\"odinger equation. In the area of turbulence modeling,
this will be the Navier-Stokes equation.  It is possible that machine learning can also play an important
role in developing the next-generation algorithms for these problems \cite{carleo2017solving, han2019solving}.

\bibliography{ref}{}
\bibliographystyle{unsrt}

\end{document}